\documentclass[12pt]{article}
\usepackage{amsfonts}
\usepackage{amssymb}
\usepackage{tikz}
\usepackage{amsmath,amssymb}
\usepackage{bbm}

\def\sfrac#1#2{{\textstyle\frac{#1}{#2}}}
\newcommand{\CR}{\mathcal{R}} 
\newcommand{\CO}{\mathcal{O}} 
\newcommand{\CH}{\mathcal{H}}  %Hilbert space
\newcommand{\R}{\mathbb{R}}     % field of real numbers
\newcommand{\C}{\mathbb{C}}     % field of complex numbers
\newcommand{\Z}{\mathbb{Z}}
\newcommand{\N}{\mathbb{N}}
\newcommand{\dd}{\mathrm{d}}     % total differential

\newcommand{\im}{\mathrm{i}} 
\newcommand{\dpar}{\partial}

\newcommand{\vp}{{\varphi}} 
\newcommand{\zb}{{\bar{z}}}
\newcommand{\zd}{{\dot{z}}}

\newcommand{\und}{{\quad\mathrm{and}\quad}}
\newcommand{\with}{{\quad\mathrm{with}\quad}}

\begin{document}
\begin{titlepage}
\setcounter{page}{0}
.
%\vskip 15mm.
\bigskip

\begin{center}
{\LARGE \bf Discrete symmetries\\[10pt]in classical and quantum oscillators}\\
\vskip 1.5cm
{\Large Alexander D. Popov}
\vskip 1cm
{\em Institut f\"{u}r Theoretische Physik,
Leibniz Universit\"{a}t Hannover\\
Appelstra{\ss}e 2, 30167 Hannover, Germany}\\
{Email: alexander.popov@itp.uni-hannover.de}\\
% {ORCID:  0000-0002-1432-495X}
\vskip 1.5cm
\end{center}
\begin{center}
{\bf Abstract}
\end{center}
We consider the nature of the wave function using the example of a harmonic oscillator.
We show that the eigenfunctions $\psi_n{=}z^n$ of the quantum Hamiltonian in the complex Bargmann-Fock-Segal representation with $z\in\mathbb C$ are the coordinates of a classical oscillator with energy $E_n=\hbar\omega n$, $n=0,1,2,...\,$. They are defined on conical spaces ${\mathbb C}/{\mathbb Z}_n$ with cone angles $2\pi/n$, which are embedded as subspaces in the phase space $\mathbb C$ of the classical oscillator. Here ${\mathbb Z}_n$ is the finite cyclic group of rotations of the space $\mathbb C$ by an angle $2\pi/n$. The superposition $\psi =\sum_n c_n\psi_n$ of the eigenfunctions $\psi_n$ arises only with incomplete knowledge of the initial data for solving the Schr\"odinger equation, when the conditions of invariance with respect to the discrete groups ${\mathbb Z}_n$ are not imposed and the general solution takes into account all possible initial data parametrized by the numbers $n\in\mathbb N$.
\end{titlepage}

\noindent
{\bf  Quantum $\longrightarrow$ classical.}  Suppose we consider a classical model with a finite-dimensional phase space $M$ and its quantum version with a Hilbert space $\CH$.  A countable basis $\{\psi_n, n=0,1,...\}$ of a complex Hilbert space $\CH$ is introduced as a system of orthogonal eigenstates of the Hamiltonian operator $\hat H$ if its spectrum is discrete and the space $\CH$ is separable. It is believed that the states $\psi_n$ from $\CH$ cannot be associated with the states of the classical model with phase space $M$. At best, when defining the connection between classical and quantum states, one considers either assigning of average values of coordinates and momenta, or embedding $M$ into $\CH$ via coherent states. In this paper we consider the classical harmonic oscillator with vector phase space $M=\C\cong\R^2$ and the space of holomorphic functions on $\C$ with a basis $\{\psi_n=z^n\}$ as the state space $\CH$ of the quantum oscillator.  We define mappings  $\CH\longrightarrow\C_n\subset\C$, showing that $\psi_n$ are coordinates on conical spaces $\C_n=\C/\Z_n$ embedded in the original phase space $\C$ of oscillator, $n\in\N$. Here $\Z_n$ is a cyclic group of rotations through an angle $2\pi/n$ in the space $\C$. The above $\psi_n$ are coordinates describing the $\Z_n$-invariant motion of classical oscillator.
Our goal in this paper is to describe this correspondence between quantum and classical states. 

\medskip

\noindent
{\bf  Classical oscillator.} We consider a classical harmonic oscillator as a particle of mass $m$ rotating in a circle $S^1$ in phase space $\R^2\cong \C$ with angular frequency $\omega$. On this phase space, parametrized by the coordinate $x$ and momentum $p$, we introduce a dimensionless complex coordinate
\begin{equation}\label{1}
z=\rho\,e^{\im\vp}=\frac{1}{r_0}\,\bigl(x-\frac{\im}{m\omega}\,p\bigr)\with r_0^2=\frac{2\hbar}{m\omega}\ ,
\end{equation}
where $r_0$ is the length parameter~\cite{Dirac}. In these coordinates, the Hamiltonian function of the oscillator has the form
\begin{equation}\label{2}
H=\frac{p^2}{2m}+\frac{m\omega^2 x^2}{2}=\hbar\omega z\zb\ ,
\end{equation}
where the Planck constant $\hbar$ appears due to its presence in the definition of $z$.

On the phase space $\C$ with coordinate  \eqref{1} we introduce the standard symplectic 2-form
\begin{equation}\label{3}
\Omega =\dd x\wedge\dd p = -\im\hbar\,\dd z\wedge\dd\zb\ ,
\end{equation}
where the bar denotes complex conjugation. The Hamiltonian \eqref{2} and 2-form \eqref{3} define a vector field 
\begin{equation}\label{4}
V_H=\Omega^{\zb z}\dpar_\zb H\dpar_z + \Omega^{z\zb}\dpar_z H\dpar_\zb=\im\omega z\dpar_z - \im\omega \zb\dpar_\zb\ ,
\end{equation}
which is the generator of the group U(1), $\dpar_z:=\dpar/\dpar z$. The equation of motion and its solution have the form 
\begin{equation}\label{5}
\zd =V_H z=\im\omega z\quad \Rightarrow\quad z(\tau)=e^{\tau V_H}z = e^{\im\omega\tau}z\ , 
\end{equation}
where the dot denotes the derivative with respect to the evolution parameter $\tau$ and  $z=z(\tau =0)$.

Comparing \eqref{1} and \eqref{5}, we see that the point moves along a circle $S^1\subset\C$ parametrized by the angle $\vp =\omega\tau$. Recall that $z$ is a dimensionless coordinate containing a length parameter $r_0$, so we choose $z(0)=1$ obtaining the conserved energy
\begin{equation}\label{6}
E=\hbar\omega
\end{equation}
as a basis for comparison with other choices of initial data.

\medskip

\noindent
{\bf  $\Z_n$-invariance.} Consider the cyclic group $\Z_n=\Z/n\Z$ of order $n$ generated by an element $\zeta\in \C$ with $\zeta^n=1$, i.e. $\zeta$ is $n$-th root of unity,
\begin{equation}\label{7}
\Z_n=\{\zeta^\ell = e^{\frac{\im2\pi\ell}{n}}, \ell =0,...,n-1\mid \zeta =e^{\frac{\im2\pi}{n}}\}\ .
\end{equation}
We define the action of this group $\Z_n\subset\,$U(1) on the coordinate $z\in\C$ by the formula
\begin{equation}\label{8}
z\ \mapsto\  \tilde z=\zeta z=\rho e^{\im(\vp + \frac{2\pi}{n})}\ ,
\end{equation}
showing that this group rotates the vector $z\in \C$ by an angle multiple of $2\pi/n$ around the point $z=0$. The Hamiltonian \eqref{2}, the symplectic 2-form \eqref{3} and the equation of motion \eqref{5} do not change under the action of this group. Therefore, we can consider the task of describing $\Z_n$-invariant solutions of the equation of motion \eqref{5} for $n\in\N$.

The solution \eqref{5} is periodic in the angular variable $\vp =\omega\tau$ with a period $2\pi$. Namely, a particle that started moving at $\tau =0$ will return to its original point after time  $\tau_1=2\pi/\omega$. Invariance under the action \eqref{8} of the group $\Z_n$ means a stronger periodicity, with the identification of points
\begin{equation}\label{9}
e^{\frac{\im2\pi}{n}}z\sim z\quad\Leftrightarrow\quad\vp +\frac{2\pi}{n}\sim\vp\ ,
\end{equation}
which means the identification of rays with the angle $2\pi/n$ between them. Here the symbol ``$\sim$" means equivalence. Therefore, with $\Z_n$-invariant motion the particle will return to the starting point in time
\begin{equation}\label{10}
\omega\tau_n=\frac{2\pi}{n}\quad\Rightarrow\quad\tau_n=\frac{2\pi}{\omega_n}\with \omega_n=\omega n\ .
\end{equation}
This is equivalent to an increase in its angular frequency and energy,
\begin{equation}\label{11}
\omega\ \mapsto\ \omega_n=\omega n\und E=\hbar\omega\ \mapsto\ E_n=\hbar\omega n\ .
\end{equation}
In order to better understand $\Z_n$-invariance \eqref{9}-\eqref{11} in the harmonic oscillator model,  we should move to coordinates in which it becomes explicit.

\medskip

\noindent
{\bf  Conical space $\C_n$.}  Let us introduce a subspace $\C_n$ of space $\C$ as part of the plane $\C$, which is cut out by rays $\vp=0$ and $\vp =2\pi/n$. The standard K\"ahler metric on $\C$ is given by $\dd s^2=\dd z\dd\zb$ and its restriction to the subspace $\C_n\subset\C$ is
\begin{equation}\label{12}
\dd {s^2}_{|\C_n}=\dd z\dd\zb_{|\C_n}=\dd\rho^2 +\frac{\rho^2}{n^2}\dd\vp_n^2 = \frac{1}{n^2}\,\rho_n^{\frac{2(1-n)}{2}}\bigl(\dd\rho_n^2 + \rho_n^2\dd\vp_n^2)\ ,
\end{equation}
where we have replaced $0\le\vp<\sfrac{2\pi}{n}$ in $z=\rho e^{\im\vp}$ with the angular variable $0\le\vp_n=n\vp<2\pi$ and $\rho_n:=\rho^n$. From \eqref{12} we see that this metric is conformally flat metric in the coordinate
\begin{equation}\label{13}
\psi_n=\rho_ne^{\im\vp_n}=z^n=\rho^ne^{\im\vp n}\ ,
\end{equation}
which is $\Z_n$-invariant coordinate on the orbifold $\C_n=\C/\Z_n$ (see e.g. \cite{RT}).

Note that from \eqref{13} we get $z=\psi_n^{1/n}$ and substituting this into \eqref{4} we get
\begin{equation}\label{14}
{V_H}_{|\C_n}=\im\omega\psi_n^{1/n}\frac{\dpar}{\dpar\psi_n^{1/n}}+c.c.=\im\omega n\psi_n\frac{\dpar}{\dpar\psi_n}-
\im\omega n\bar\psi_n\frac{\dpar}{\dpar\bar\psi_n}\ .
\end{equation}
Accordingly, the equation of motion \eqref{5} becomes the equation
\begin{equation}\label{15}
\dot\psi_n=\im\omega_n\psi_n\quad\Rightarrow\quad\psi_n(\tau)=e^{\im\omega_n\tau}\psi_n\with \omega _n=\omega n\ .
\end{equation}
It is easy to see that this equation follows not only from \eqref{2}-\eqref{5} together with \eqref{13}, but also from the Hamiltonian $H_n$ defined as
\begin{equation}\label{16}
\frac{H_n}{\hbar\omega_n}:=\left(\frac{H}{\hbar\omega}\right)^n=z^n\zb^n=\psi_n\bar\psi_n\quad\Rightarrow\quad
H_n=\hbar \omega_n\psi_n\bar\psi_n\ ,
\end{equation}
and the vector field \eqref{14} derived from \eqref{16} with the symplectic structure $\Omega_n$ on $\C_n$ of the form
\begin{equation}\label{17}
\Omega_n=-\im\hbar\,\dd\psi_n\wedge\dd\bar\psi_n\ .
\end{equation}
This 2-form differs from $\Omega_{|\C_n}$ by the same conformal factor that the flat metric $\dd\psi_n\dd\bar\psi_n$ differs from the metric \eqref{12}.

The Riemann curvature of the metric \eqref{12} is
\begin{equation}\label{18}
\CR =\frac{2\pi (n-1)}{n}\delta(\varrho_n)\,\dd\psi_n\wedge\dd\bar\psi_n\ ,
\end{equation}
i.e. it is flat everywhere except the fixed point $\psi_n=0=z$. Note that this point is excluded when considering quantum oscillator (see e.g. \cite{Sni}). Recall that $z$ and $\psi_n$ are dimensionless complex coordinates, and the energy of $\Z_n$-invariant oscillator with the choice $z\zb =1$ from \eqref{6} take the form
\begin{equation}\label{19}
E_n=\hbar\omega n\ ,
\end{equation}
i.e. it increases by $n$ times.

\medskip

\noindent
{\bf  Covering map.}  In the language of differential geometry, the transition from the space $\C$ to its $\sfrac1n$-part $\C_n$ is carried out through the equivalence relation \eqref{9} given by the action \eqref{8} of the group $\Z_n$ and the associated covering  map. A covering is a map between topological spaces  $\tilde X$ and $X$ that locally acts like a projection of a multiple copies of a space $X$ onto itself.  In particular, the mapping
\begin{equation}\label{20}
\psi_n:\quad\C\stackrel{\Z_n}{\longrightarrow}\C/\Z_n\with \psi_n(z)=z^n
\end{equation}
is a {\it branched} covering of degree $n$, where $z=0$ is the branch point for $n\ge 2$. The function $\psi_n(z)$ in \eqref{13}-\eqref{20} is an ordinary holomorphic function and its inverse is a multivalued function corresponding to the map
\begin{equation}\label{21}
\psi_n^{-1}:\quad\C/\Z_n\longrightarrow\C\ , \quad z=\psi_n^{1/n}\ ,
\end{equation}
embedding $\C/\Z_n$ into the space $\C$ as part $\C_n$ of the plane $\C$ described in \eqref{12}-\eqref{18}.

To summarize, the description of a $\Z_n$-invariant oscillator \eqref{1}-\eqref{11} on the plane $\C$ corresponds to the transition  \eqref{12}-\eqref{20} to a cone $\C/\Z_n=C(S^1/\Z_n)$ with a cone angle $2\pi/n$ and coordinate $\psi_n$:
\begin{center}
\begin{tikzpicture}
\draw (0,0) -- (-1,-2);
\draw (0,0) -- (1,-2);
\filldraw[black] (0,0) circle (2pt);
\node at  (0.7,0) {$z=0$};
\node at  (0,-1.8) {$\frac{2\pi}{n}$};
\node at  (0,-1) {$\C_n$};
\node at  (2,-1) {$\C, z$};
\draw[->, thick] (2.8,-1) -- (3.8,-1); 
\draw (6,0) -- (5,-2);
\draw (6,0) -- (7,-2);
\filldraw[black] (6,0) circle (2pt);
\node at  (7,0) {$\psi_n=0$};
\node at  (8.1,-1) {$\C_n, \psi_n=z^n$};
%\draw (6, -2) ellipse (1 and 0.25);
\draw (5,-2) .. controls (5.5,-2.5) and (6.5,-2.5) .. (7,-2);
\draw[dashed](5,-2) .. controls (5.5,-1.5) and (6.5,-1.5) .. (7,-2);
\end{tikzpicture}
\end{center}
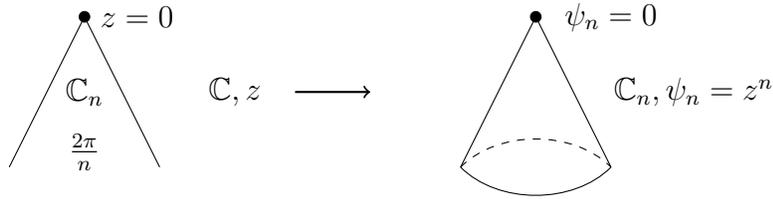
{\footnotesize{\bf Figure 1.} The interior of the angle $\sfrac{2\pi}{n}$ on the left is the preimage of cone $\C_n=\C/\Z_n$ in $\C$. The angle $0\le\vp <\sfrac{2\pi}{n}$ in $\C$ corresponds to the angle $0\le\vp_n <2\pi$ in $\C/\Z_n$ with $\vp_n=n\vp$.}

\medskip

\noindent Note that on $\C\backslash\{0\}$ there are $n$ different points $z_\ell =\zeta^\ell z\in\C\backslash\{0\}$ mapped to the same point $\psi_n(z_\ell)=z^n$ for $\ell =1,...,n$ (covering). In further consideration we will also include the function $\psi_0(z) =z^0=1$ corresponding to the space $\C_0=\C/\Z_0=\{$point $1\in\C\}.$

\bigskip

\noindent
{\bf  Quantum oscillator.}  Let us now turn to the quantum harmonic oscillator in the complex Bargmann-Fock-Segal representation, which uses the coordinate $z$ on the phase space $\C$ \cite{Dirac, Bar, Segal, Hall}.  In this representation, the quantum oscillator has as its state space the Hilbert space $\CH$ of holomorphic functions on $\C$ square-integrable with Gaussian weight $\exp(-z\zb)$. Inner product on $\CH$ is given by the integral
\begin{equation}\label{22}
\langle\psi , \phi\rangle=\int_\C\bar\psi (z)\phi(z)\,e^{-z\zb}\dd v\ ,
\end{equation}
where $\dd v = \rho\,\dd\rho\dd\vp$ for $z=\rho\, e^{\im\vp}$. An orthogonal basis in complex vector space $\CH$ of holomorphic functions $\psi$ with finite norm $\|\psi\|=\langle\psi , \psi\rangle^{1/2}$ are functions
\begin{equation}\label{23}
\psi_n(z)=z^n\ ,
\end{equation}
whose squared norm is equal $\|\psi_n\|^2=n!$. 

The Hamiltonian function \eqref{2} leads to an operator
\begin{equation}\label{24}
\hat H=\hbar\omega z\dpar_z + \sfrac12\hbar\omega\ ,
\end{equation}
where the zero-point energy $\sfrac12\hbar\omega$ arises from the non-zero curvature of the complex line bundle over phase space $\C$ (see e.g. \cite{Popov1}). The functions \eqref{23} are eigenvectors of the operator \eqref{24} with energy eigenvalues
\begin{equation}\label{25}
\tilde E_n=E_n+\sfrac12\hbar\omega\with E_n=\hbar\omega n\ .
\end{equation}
Any vector $\psi$ from $\CH$ can be expanded in basis \eqref{23},
\begin{equation}\label{26}
\psi = \sum\limits_{n=0}^{\infty}\frac{1}{\sqrt{n!}}\, c_n\psi_n\quad\in\quad \CH=\mathop\oplus\limits_{n=0}^{\infty}\C\ ,
\end{equation}
where $c_n$ are complex numbers. In the standard interpretation, if the vector $\psi$ from \eqref{26} is normalized to unity, then it defines a state for which the probability that the oscillator energy when measured will have the value \eqref{25} is equal to $|c_n|^2$. If we act on $\psi$ with the operator $\exp(\sfrac\im\hbar\,\tau \hat H)$, then all $\psi_n$ in \eqref{26} will receive a dependence on $\tau$ of the form \eqref{15} times $\exp(\sfrac{\im}{2}\,\omega\tau)$ due to zero-point energy. Note that we have replaced $t\mapsto-\tau$ in the Schr\"odinger equation and in the definition of positive frequency to better match the definitions of differential  geometry and group representation theory.

\bigskip

\noindent
{\bf  Classical vs quantum.}  Comparing the descriptions of classical and quantum oscillators, we see that the eigenfunctions $\psi_n$ of the quantum Hamiltonian $\hat H$ are $\Z_n$-invariant solutions of the classical oscillator equation. In this case, they are coordinates $\psi_n$ on phase spaces, nested one inside the other, for $\Z_n$-periodic oscillators
\begin{equation}\label{27}
\C =\C_1\supset\C_2\supset\ldots\supset\C_n\supset\dots \ ,
\end{equation}
in which the oscillating particle moves along circles
\begin{equation}\label{28}
S^1/\Z_n\subset\C_n=\C/\Z_n\ ,\ n=1,2,\dots\ . 
\end{equation}
Therefore, the basis vectors of the Hilbert space $\CH$ of a quantum oscillator can be identified with  $\psi_n\in\C_n$ describing $\Z_n$-invariant classical oscillator, and hence
\begin{equation}\label{29}
\mathop\oplus\limits_{n=0}^\infty\C_n=\C_0\oplus\C/\Z_1 \oplus\C/\Z_2\oplus\dots\oplus\C/\Z_n\oplus\dots\subset \CH\ , 
\end{equation}
where $\C_0=\{$point$\}=1$ (vacuum, no rotation in $\C$) and $\C/\Z_1=\C$, $\Z_1=\,$Id.

All the formulae given earlier indicate that the particle moves in a circle $S_n^1=\{0\le{\vp_n}<{2\pi}\}$ in {\it one} of the subspaces $\C_n$ of the Hilbert space \eqref{29} and has an energy consisting of the energy $E_n$ of its rotation in $\C/\Z_n$ and the energy $E_0=\sfrac12\hbar\omega$ of rotation of the vacuum vector. The fixation of the subspace $\C_n\subset\C$ at the level of the Schr\"odinger equation is determined by the requirements of invariance of the function $\psi(z)$ with respect to one or another cyclic group (conditions  of $\Z_n$-periodicity). For example, one can require that the function $\psi(z)$ be invariant with respect to the action of the group $\Z_n$ and non-invariant with respect to the action of the groups $\Z_m$ with $m>n$. 

Thus, the state $\psi_n\in\CH$ of the quantum oscillator corresponds to the state of a classical particle moving along a circle $S^1/\Z_n$ in the cone $\C/\Z_n=C(S^1/\Z_n)$, $n\in\N$. This confirms 't~Hooft's idea \cite{Hooft} that the wave function \eqref{26} reflects the probabilities $|c_n|^2$ of particle motion in one of the subspaces $\C_n\subset\C\subset\CH$ if the initial state and its symmetries are not known with absolute accuracy. There is no need to go to discrete space and time, as suggested by 't~Hooft \cite{Hooft}, since discreteness is associated with representations
\begin{equation}\label{30}
\phi_n:\ e^{\im\vp}\to e^{\im n\vp}\und \psi_n:\ z\to z^n
\end{equation}
of groups U(1) and GL(1, $\C$) and coverings
\begin{equation}\label{31}
\phi_n:\ S^1\stackrel{\Z_n}{\longrightarrow} S^1/\Z_n\und \psi_n:\ \C\stackrel{\Z_n}{\longrightarrow} \C/\Z_n\ ,
\end{equation}
where $n$ is the winding number. We conclude that the wave function  \eqref{26} describes the possible motion of classical particle and probabilities $|c_n|^2$ reflect the degree of our ignorance of the initial conditions and symmetries imposed on the motion of the particle.  In other words, when we want to measure energy, the wave function $\psi$ carries information about possible energies of classical particle, which corresponds to the epistemic view. This does not mean that we know all the other parameters of the particle's motion.

\bigskip

\noindent
{\bf  General classical solution.}  We have shown that the Hilbert space of wave functions $\CH$ from \eqref{26}-\eqref{31} describes a countable set of states of the classical oscillator. It is obvious that the classical oscillator has many more states than the set $\{\psi_n\}$, $n\in\N$. First, for the standard oscillator \eqref{1}-\eqref{5} the energy can take any positive value
\begin{equation}\label{32}
E_\gamma =\hbar\omega z\zb =\hbar\omega\gamma\ ,
\end{equation}
where $z=z(\tau =0)$ is given by the initial values of the coordinate $x(0)$ and momentum $p(0)$ of the particle. This corresponds to the particle moving along a circle of radius $\sqrt{\gamma}r_0$ in $\C$ and not of radius $r_0$, as chosen in \eqref{6}.

Note that the energy of a quantum oscillator increases due to a change in the angular frequency $\omega\mapsto\omega_n=\omega n$ and not due to a change in the initial data for the coordinate and momentum. As we discussed in detail, this change in energy of a classical particle is related with $\Z_n$-invariance. In this case, the particle moves in a conical space $\C_n\subset\C$ with a cone angle $2\pi/n$. However, the conical space $\C_\gamma$ can be defined for any angle $2\pi/\gamma$, which also leads to a classical state with energy  \eqref{32}, where  the change in energy is associated  with a change in the angular frequency $\omega\mapsto \omega\gamma$, and not with a change in the initial data $z(0)$. 

To clarify the above, let us first consider $\gamma\ge 1$ and introduce a subspace $\C_\gamma$ in $\C$ replacing the space $\C_n$ in Fig.1. We restrict the metric on $\C$ to this subspace,
\begin{equation}\label{33}
\dd s^2_{|\C_\gamma}=\dd\rho^2 + \frac{\rho^2}{\gamma^2}\,\dd\vp^2_\gamma =
\frac{1}{\gamma^2}\rho_\gamma^{2(1-\gamma)/\gamma}\bigl(\dd\rho_\gamma^2 +
\rho_\gamma^2\dd\vp_\gamma^2\bigr)\ ,
\end{equation}
where $0\le\vp_\gamma =\gamma\vp<2\pi$ and $\rho_\gamma:=\rho^\gamma$, similar to formula \eqref{12}. We obtain a conformally flat space $\C_\gamma$ with coordinate
\begin{equation}\label{34}
\psi_\gamma =\rho_\gamma e^{\im\vp_\gamma} =z^\gamma\ .
\end{equation}
Conical spaces with metric of the form \eqref{33} have been used in applications to many problems in physics (see e.g. \cite{SS, BV, Bordag}).

For $\gamma\ge 1$, the cone in Fig.1 is preserved with $n$ replaced by $\gamma$, but the cone $\C_\gamma$ is no longer a coset space. The difference from the case $\gamma =n$ is that the mapping
\begin{equation}\label{35}
\psi_\gamma :\ \C \longrightarrow \C_\gamma\with \psi_\gamma (z) =z^\gamma
\end{equation}
is defined by a multivalued function. When choosing a specific branch, the space $\C_\gamma$ can be considered as a cone over a circle $S_\gamma^1$ along which a particle moves with
\begin{equation}\label{36}
\psi_\gamma (\tau) =\exp(\im\omega\gamma\tau)\psi_\gamma(0)\ .
\end{equation}
Note that the value $0<\gamma <1$ can also be considered in all formulae and the only difference is that the angle $2\pi/\gamma$ will be greater than $2\pi$ (origami-style cone).

Note that $\psi_\gamma = z^\gamma$ is a solution to the Schr\"odinger  equation
\begin{equation}\label{37}
\hat H\psi_\gamma =\tilde E_\gamma\psi_\gamma\with \tilde E_\gamma =\hbar\omega (\gamma + \sfrac12).
\end{equation}
Thus, all classical states of the harmonic oscillator with $\gamma >0$ are eigenfunctions of the quantum Hamiltonian. The difference between function $\psi_\gamma$ and $\psi_n$ is that $\psi_\gamma$ is multi-valued for $\gamma\ne n$ and is not holomorphic function on the complex plane $\C$. Therefore, the functions $\psi_\gamma$ do not belong to the Hilbert space $\CH$ of the quantum oscillator described in \eqref{22}-\eqref{29}. Functions \eqref{35} have a branch point at $z=0$ and the value of
\begin{equation}\label{38}
\psi_\gamma =z^\gamma =\exp (\gamma\log z)
\end{equation}
depends on the branch of the logarithm used. One must choose a specific branch by defining a branch cut (usually the positive real axis) along which the function is discontinuous. Thus, these eigenfunctions of the Hamiltonian are unacceptable from the point of view of quantum mechanics, but they are normal as coordinates of a classical oscillator.  Note that in the position representation the situation is similar (see e.g. \cite{Fermi}), where $\psi_\gamma$ with $\gamma\ne n$ grows as an exponential function $\exp(x^2/r_0^2)$ as $x^2$ tends to infinity. In quantum mechanics, such solutions are rejected by the requirement of belonging to a Hilbert space.

\bigskip

\noindent
{\bf  Concluding remarks.}  We have shown that the wave function \eqref{26} of a quantum harmonic oscillator is a formal sum of coordinates $\psi_n$ on phase spaces $\C/\Z_n$ of a classical oscillator with $\Z_n$-invariance.  This oscillator is described by a point moving along a circle $S^1/\Z_n\subset\C/\Z_n$, given by formula \eqref{15}, obtained by transforming the coordinates $z\mapsto\psi_n(z)=z^n$  on the plane $\C$. The cyclic group $\Z_n$ is related to one-dimensional complex representations \eqref{30} of the group U(1) and GL(1, $\C$), winding numbers $n\in\pi_1(S^1)$ and the concept of periodicity.

Let us recall how the concept of periodicity is introduced. The real line $\R$ is divided into an infinite number of intervals,
\begin{equation}\label{39}
\R = \mathop\coprod\limits_{k=-\infty}^{\infty}\bigl[2\pi k, 2\pi (k+1)\bigr)\ \stackrel{\Z}{\longrightarrow}\ \R/\Z =S^1\ ,
\end{equation}
where $\Z$ is a group of integers. The group $\Z$ of shifts on $2\pi$ defines the equivalence of points on different intervals in \eqref{39}, that  is why the circle $S^1$ is identified with the segment $[0, 2\pi]$, in which the points $x=0$ and $x=2\pi$ are identified, and then $S^1$ is parametrized by the angular variable $0\le\vp <2\pi$.

The group $\Z_n$ is the group of integers modulo $n$ and its use allow us to strengthen the periodicity $f(x+2\pi)=f(x)$ of function on $\R$ to periodicity on a circle $S^1$. To do this, we divide the circle $S^1=[0, 2\pi)$ into $n$ intervals,
\begin{equation}\label{40}
S^1 = \mathop\coprod\limits_{\ell=0}^{n-1}\left[\frac{2\pi \ell}{n}, \frac{2\pi (\ell+1)}{n}\right)\ \stackrel{\Z_n}{\longrightarrow}\ S^1/\Z_n =:S^1_n\ ,
\end{equation}
where the group $\Z_n$ of shifts $\vp\mapsto\vp + \sfrac{2\pi}{n}$ defines the equivalence of points of different intervals in \eqref{40}. That is why the circle $S_n^1$ with the angular variable $0\le\vp_n <2\pi$ can be identified with the segment $[0, 2\pi/n]\subset S^1$, in which the points $\vp =0$ and $\vp =2\pi/n$ coincide. The projection \eqref{40} induces the projection $\C\longrightarrow\C/\Z_n$ and therefore the space $\C/\Z_n$ can be identified with the subspace $\C_n$ in $\C$ bounded by the rays $\vp=0$ and $\vp=2\pi/n$. Accordingly, the condition of $\Z_n$-invariance of a function $\psi(z)$ of $z\in\C$ is the periodicity condition $\psi(\zeta z) = \psi (z)\,\Rightarrow\,\psi (\rho, \vp + 2\pi/n)=\psi(\rho, \vp)$ for $\zeta =\exp(\im 2\pi/n)\in\Z_n$.

We emphasize that we do not use any interpretations of quantum mechanics and do not propose new ones. We simply view the oscillator through the prism of differential geometry and do not see sharp boundary between the classical and quantum cases. In the ``quantum" case, the energy operator $\hat H$ acts on vectors $\psi$ from the Hilbert space $\CH$ of holomorphic functions on $\C$ with basis $\psi_n=z^n$ (eigenfunctions),
\begin{equation}\label{41}
\psi (z) =\sum_{n=0}^\infty\frac{1}{\sqrt{n!}}\,c_n\psi_n\ .
\end{equation}
The dynamics are introduced by the action of the evolution operator on $\psi\in\CH$,
\begin{equation}\label{42}
\psi (z, \tau) =\exp\bigl(\sfrac{\im}{\hbar}\tau\hat H\bigr)\psi =\exp\bigl(\sfrac{\im\omega\tau}{2}\bigr)\sum_{n=0}^\infty\frac{1}{\sqrt{n!}}\,c_n\psi_n(\tau)\ ,
\end{equation}
where 
\begin{equation}\label{43}
\psi_n(\tau) = e^{\im\omega n\tau}\psi_n =e^{\im\omega n\tau}z^n\ .
\end{equation}
In fact, we have an infinite-dimensional representation of the group U(1) on the direct sum \eqref{29} of spaces of one-dimensional irreducible representations \eqref{30}, \eqref{31}. We have shown that the states \eqref{43} of a quantum oscillator are coordinates on the conical phase spaces $\C_n=\C/\Z_n$ of a classical $\Z_n$-invariant oscillator.

Imposing the $\Z_n$-invariance condition before choosing a solution at the classical and quantum level  turns energy into an ontological variable before any evolution and the collapse $\psi(z)\to\psi_n(z)$ loses its meaning. In this case, superposition \eqref{41} reflects incomplete knowledge about which of the subspaces $\C_n$ of phase space $\C$ the particle is moving in. Accordingly, one can repeat after 't~Hooft \cite{Hooft} that the theory reproduces the probability distribution of final states through the probability distribution of initial states. Let me make it clear that we are talking only about energy - I do not want to make any statements about other observables (coordinate, phase, etc.).

Recall that the number $n\in\pi_1(S^1)$ in $\Z_n$ is the winding number introduced in \eqref{30}. Negative values $k=-n$ of this number can be obtained by considering $\psi_{-n}=z^{-n}$ or better $\psi_{-n}=\zb^{n}$ together with the changed sign of the symplectic structure \cite{Popov1}. In fact, topological numbers of this type have a common mathematical nature. Namely, let us recall that holomorphic line bundles $\CO_N(k)$ over a complex projective space $\C P^N$ are defined by an equivalence relation
\begin{equation}\label{44}
\Bigl([\lambda z_0:...:\lambda z_N], \lambda^k\psi_k\Bigr)\sim\Bigl([z_0:...:z_N], \psi_k\Bigr)\in\CO_N(k)\ ,
\end{equation}
where $[z_0\!:...:\!z_N]$ are homogeneous coordinates on $\C P^N$, $\lambda\in\,$GL(1, $\C$), $k\in\Z$ and $\psi_k\in\C$ is a component of a complex {\it vector} in the fibre, i.e. linearity and the possibility of superposition in this case are already defined at the classical level (see e.g. \cite{Wells}). The numbers $k=\pm n\in\Z$ here are the Chern classes of the bundle $\CO_N(k)$. The fibres in the neighborhood of zero section of the bundle $\CO_N(-n)$ are the orbifolds $\C/\Z_n$ considered in this paper.
In the paper \cite{Popov2} we discussed in detail the case $N=1$, which describes the spin of particles. The consideration of this paper  formally corresponds to the case $N=0$, in which Chern classes are replaced by winding numbers. The dynamics are determined by restricting the parameter $\lambda\in\,$GL(1, $\C$) to the subgroup U(1)$\subset$GL(1, $\C$), which gives the rotation $\psi_k(\tau)=\exp (\im\omega k\tau)\psi_k(0)$ of the particle in one-dimensional fibres of the bundle \eqref{44} over the space $\C P^N$ parametrizing the initial data of this motion.

\bigskip

\noindent
{\bf  Measurement problem.} The measurement problem in quantum mechanics is the fundamental question of how a quantum system, existing in a superposition  of states, chooses one specific outcome when measured.  This problem and the related problem of interpreting quantum mechanics have remained unsolved since the creation of quantum mechanics (see e.g. \cite{S1, S2, Tomaz} for discussion and references). In fact, the problem arises because we introduce into a mathematically formalized theory (Hilbert spaces, operators, and so on), which claims to be fundamental, such elements external to it as ``device", ``environment", ``observer", and others. Quantum information theory improves the situation by replacing these words with formally defined objects and rules for operating on them. However, all this is external to quantum mechanics and therefore does not solve the measurement problem - only states it cleanly.

In this paper we have considered the question of the nature of wave function as a superposition of states, using methods  of 
differential geometry applied to the harmonic oscillator. We have shown that the basis vectors $\psi_n$ of the Hilbert space $\CH$ of holomorphic functions \eqref{41} are coordinates on the conical phase spaces $\C_n=\C/\Z_n\subset\C$ of $\Z_n$-invariant classical harmonic oscillator. We have established the following correspondences:
\bigskip
\begin{tabbing}
nnn\=\hspace{8cm}\=\hspace{8cm}   \kill
$~~~~~~$classical\>\>$~~~~~~$quantum\\[3pt]
$\psi_n{\in}\C_n{\subset}\C,\ E_n=\hbar\omega n, n\in\Z_{\ge 0}$\>\>$\psi{=}\sum\limits_{n=0}^\infty\frac{1}{\sqrt{n!}} c_n\psi_n{\in}\CH,\ \tilde E_n=E_n+\sfrac12\hbar\omega$\\[3pt]
$\psi_\gamma{\in}\C_\gamma{\subset}\C,\ E_\gamma{=}\hbar\omega\gamma , \gamma\in\R^+, \gamma\ne n$\>\>eigenstate $\psi_\gamma$ of $\hat H$ outside $\CH$
\end{tabbing}
\smallskip
Our description shows that the classical states $\psi_n$ form a basis of the Hilbert space $\CH$ and therefore the wave function $\psi$ of the ``quantum" oscillator is not a physical field. The superposition $\psi$ of states $\psi_n$ defines the probability distribution (via $c_n$) of the choice of $\Z_n$-invariant motion of the classical oscillator, that is, the wave function defines a statistical description. Let us note that this is not a hypothesis or an interpretation, but a mathematical fact.

%\newpage

\end{document}